\documentclass[aps,prl,showpacs]{revtex4}
\usepackage{graphicx}

\begin{document}
\title{The Fermi surface of Na$_{x}$CoO$_2$}

\author{Peihong Zhang}
\author{Weidong Luo}
\author{Marvin L. Cohen}
\author{Steven G. Louie}

\affiliation{Department of Physics, University of California at Berkeley, Berkeley, California 94720}
\affiliation{Materials Sciences Division, Lawrence Berkeley National Laboratory, Berkeley, California 94720}

\date{\today}

\begin{abstract}
Doping evolution of the Fermi surface topology of Na$_x$CoO$_2$ is studied 
systematically. Both local density approximation (LDA) and local spin density
approximation (LSDA) predict a large Fermi surface as well as small hole pockets 
for doping levels $x\sim$ 0.5. In contrast, the hole pockets are completely
absent for all doping levels within LSDA+U. More importantly, we find no violation 
of Luttinger's rule in this system, contrary to a recent suggestion. The measured 
Fermi surface of Na$_{0.7}$CoO$_2$ can be explained by its 
half-metallic behavior and agrees with our LSDA+U calculations.
\end{abstract}

\maketitle

Detailed knowledge of the Fermi surface topology is important for understanding 
various properties of metals. In the conventional BCS theory of superconductivity, 
electron-phonon 
couplings, and hence the superconducting transition temperature, depend sensitively 
on the electronic structure near the Fermi surface. In most unconventional 
superconductivity scenarios, the low energy excitations and pairing 
interactions of the system are presumably also determined by the electronic structure 
near the Fermi surface. Therefore, an accurate description of the Fermi surface 
is critical for both types of superconductivity.
Recently, the discovery of superconductivity\cite{supercond} in hydrated Na$_x$CoO$_2$
has generated renewed interest in this material and much effort has focused on the 
understanding of its normal state electronic 
properties.\cite{Lee04,Zhang04,Marianetti04,Johannes04_1,Li04,Johannes04_3,Wu04,Zou04} 
One central issue 
is the shape (and the size) of the Fermi surface. Earlier  
LDA calculations\cite{NaCo2O4} of the unhydrated system (Na$_{0.5}$CoO$_2$) 
suggested a rather interesting picture: A large circular Fermi surface surrounded 
by small pockets of holes near the K points in the Brillouin zone (BZ). This Fermi 
surface topology is the starting point for several proposed theories of 
non-phonon mediated
superconductivity\cite{Kuroki04,Johannes04_2}, and the existence of small hole pockets 
are essential for these theories. However, recent angle-resolved photoelectron 
spectroscopy (ARPES) experiments by at least two groups revealed a much simpler Fermi 
surface.\cite{Hasan04,Yang04} More surprisingly, the measured Fermi surface 
does not seem to satisfy the Luttinger electron counting rule if a half-filled 
paramagnetic two-dimensional band structure is assumed for the undoped system (CoO$_2$).\cite{Hasan04}

In this paper, the Fermi surface of Na$_x$CoO$_2$ ($x=0.3$, 0.5, and 0.7) is calculated 
with three levels of approximations, namely, LDA, LSDA, 
and LSDA+U\cite{LDA+U} with a moderate U of 4.0 eV. The system is modeled with a single 
layer CoO$_2$ doped with electrons and we assume that no charge disproportionation 
occurs in the metallic phase. Charge neutrality is assured by a balancing uniform 
positive background. We believe that this model captures the essential physics,  
and the presence of the Na potential will have minor effects on our results.\cite{Zhang04}
Charge disproportionation\cite{Foo04,Li04,Mukhamedshin04,Huang04,Wang04,Lee042} 
and its implications on the observed insulating phase of 
Na$_{0.5}$CoO$_2$ at low temperature,\cite{Foo04} as well as the subtle interplay between the Na 
ordering and the charge ordering in the CoO$_2$ layer, will be discussed in a separate 
publication. We find that the measured Fermi surface\cite{Hasan04} of Na$_{0.7}$CoO$_2$
can well be explained by the half-metallic behavior of the system. Therefore,
contrary to a previous suggestion,\cite{Hasan04} the Luttinger rule is obeyed.

Figure 1 shows the LDA band structures of (CoO$_2$)$^{x-}$ ($x=$0.3, 0.5, and 0.7) 
near the Fermi level. Only the Co $d$ derived $t_{2g}$ states are shown for clarity. 
Under the influence of the trigonal crystal field, the $t_{2g}$ 
triplet splits into an $a_{1g}$ state and an $e_{g}$ doublet with a separation of
about 1.5 eV at $\Gamma$. Interestingly, the bandwidth of the $e_{g}$ doublet is 
also about 1.5 eV within LDA. As a result, the Fermi level crosses both the $a_{1g}$ and one of
the $e_{g}$ bands at low doping level ($x\le 0.6$), producing a large Fermi 
surface around the $\Gamma$ point as well as small hole pockets near the K points in 
the BZ (see also Fig. 4). This is in good agreement with the result of Singh.\cite{NaCo2O4}
The small hole pockets disappear as the 
doping level $x$ increases beyond 0.6, which seems to agree with the fact that no hole
pockets are observed\cite{Hasan04,Yang04} in Na$_{0.6}$CoO$_2$ and Na$_{0.7}$CoO$_2$. However, 
due to a dip in the band energy of the $a_{1g}$ state at $\Gamma$, a feature that can be understood 
within a second-nearest-neighbor tight-binding model, the Fermi surface
splits into two concentric pieces (see Fig. 4 for details) at $x=0.7$. This is not
consistent with experiments where only one cylindrical Fermi surface
was observed.\cite{Hasan04} Moreover, the radius of the Fermi surface calculated within
LDA is considerably smaller than the observed values,\cite{Hasan04} which raised the
issue of possible violations of the Luttinger rule. However, there
exists a simpler explanation, as will be discussed in the following.

Considering the relatively narrow bandwidth of the $d$ states and the observed
magnetism\cite{magnetism1,magnetism2} in this system, one might expect that 
spin-polarized calculations within LSDA be more appropriate for describing its
electronic structure. The calculated LSDA band structures of (CoO$_2$)$^{x-}$ 
near the Fermi level are shown in Fig.\ 2. The spin polarization energy pulls the majority
spin $a_{1g}$ state down by about 0.61, 0.49, and 0.34 eV at the $\Gamma$ point for 
doping levels $x=$ 0.3, 0.5, and 0.7, respectively. As a result, the $t_{2g}$ derived
majority spin states lie completely below the Fermi level and the system becomes a half-metal.
This is in agreement with a previous LSDA calculation.\cite{NaCo2O4} At 
low doping level ($x=0.3$), the Fermi level cuts all three sub-bands of the 
minority spin $t_{2g}$ triplet, resulting in a rather complicated Fermi surface, 
as shown in Fig.\ 4. As the doping level $x$
increases to 0.5, only two of the three sub-bands cross the Fermi level
and hole pockets, considerably larger than those calculated within LDA, develop near 
the K points. At $x=0.7$, only the $a_{1g}$ state contribute to the Fermi surface 
and the small hole pockets disappear. Due to its half-metallic behavior, the phase
space volume (holes) enclosed by the Fermi surface is twice as large as it would be
if both spins contribute. Our calculation gives an average radius of $k_f \sim 0.72$ \AA$^{-1}$
for $x=0.7$, which is in reasonable agreement with the measured value\cite{Hasan04}  
(0.65 $\pm 0.1$ \AA$^{-1}$ for $x=0.7$). Therefore, we conclude that the half-metallic
behavior of the system is consistent with the measured Fermi surface topology
and that Luttinger's rule is satisfied.

Although there is still no consensus on the strength of the on-site Coulomb
interaction $U$ among the Co $d$ electrons in the CoO$_2$ plane, it is generally 
agreed that Na$_x$CoO$_2$ is a moderately correlated system. (It is also possible
that the correlation effects vary with the doping level due to the change
in electron screening.\cite{Lee042}) Therefore, the applicability of LDA or LSDA 
may be questionable. Here we employ the LSDA+U
method introduced by Anisimov {\it et al.}\cite{LDA+U} to study the electronic band structure
and the Fermi surface of this system. The LSDA+U method can be regarded as a model 
Hartree-Fock approximation with the Coulomb interactions among the localized (e.g., Co $d$) 
electrons being replaced by statically screened parameters $U$ and $J$. In addition 
to the splitting arising from 
trigonal crystal field effects, the on-site Coulomb interaction further pushes the 
partially occupied minority spin $a_{1g}$ state up, resulting in a complete
separation between the $a_{1g}$ and the $e_{g}$ states at low doping levels, as shown
in Fig.\ 3. Consequently, the small hole pockets are absent for all doping
levels within LSDA+U, as shown in Fig.\ 4. In fact, a very small on-site Coulomb
$U$ ($\sim 2$ eV) is sufficient to suppress the hole pockets at low doping levels.
Therefore, whether the hole pockets exist or not is a very subtle issue, especially 
at low doping levels. The absence of small hole pockets, if further confirmed,
will strongly favor an intermediate correlation picture. We hope our results will 
stimulate more experimental work on this subject.

Finally, we compare the calculated Fermi surface within LDA, LSDA, LSDA+U with that
measured by ARPES experiments\cite{Hasan04}, as shown in Fig.\ 5. Both LSDA and LSDA+U
calculations give a Fermi surface which agrees well with the experiment, indicating
that Luttinger's rule is satisfied. LDA, on the other hand, results in a qualitatively
different Fermi surface topology from the measured one. 

In summary, we have studied the electronic band structure and the Fermi surface topology
of Na$_x$CoO$_2$ using three levels of approximations, namely, LDA, LSDA, and LSDA+U 
with a moderate Coulomb $U$ of 4 eV. The calculated Fermi surface structure is nontrivial
and depends sensitively on the doping level as well as on the theoretical approximations.
Both LDA and LSDA predict a large Fermi surface around the $\Gamma$ point
and small hole pockets near the K points at doping levels $x\le 0.5$.
In contrast, no hole pockets are observed within LSDA+U for all doping levels.
The measured Fermi surface at doping $x=0.7$ agrees well with our LSDA and LSDA+U
calculations, and we find no violation of Luttinger's rule in this
system.

\begin{acknowledgments}
This work was supported by National Science Foundation Grant No.\ DMR-0087088
through the Center of Materials Simulation and Office 
of Energy Research, Office of Basic Energy Sciences, 
Materials Sciences Division of the U.S. Department of Energy under Contract 
No.\ DE-AC03-76SF00098. Computational Resources were provided by NPACI and NERSC.
\end{acknowledgments}

\newpage

\begin{figure}[h]
\includegraphics[width=8cm]{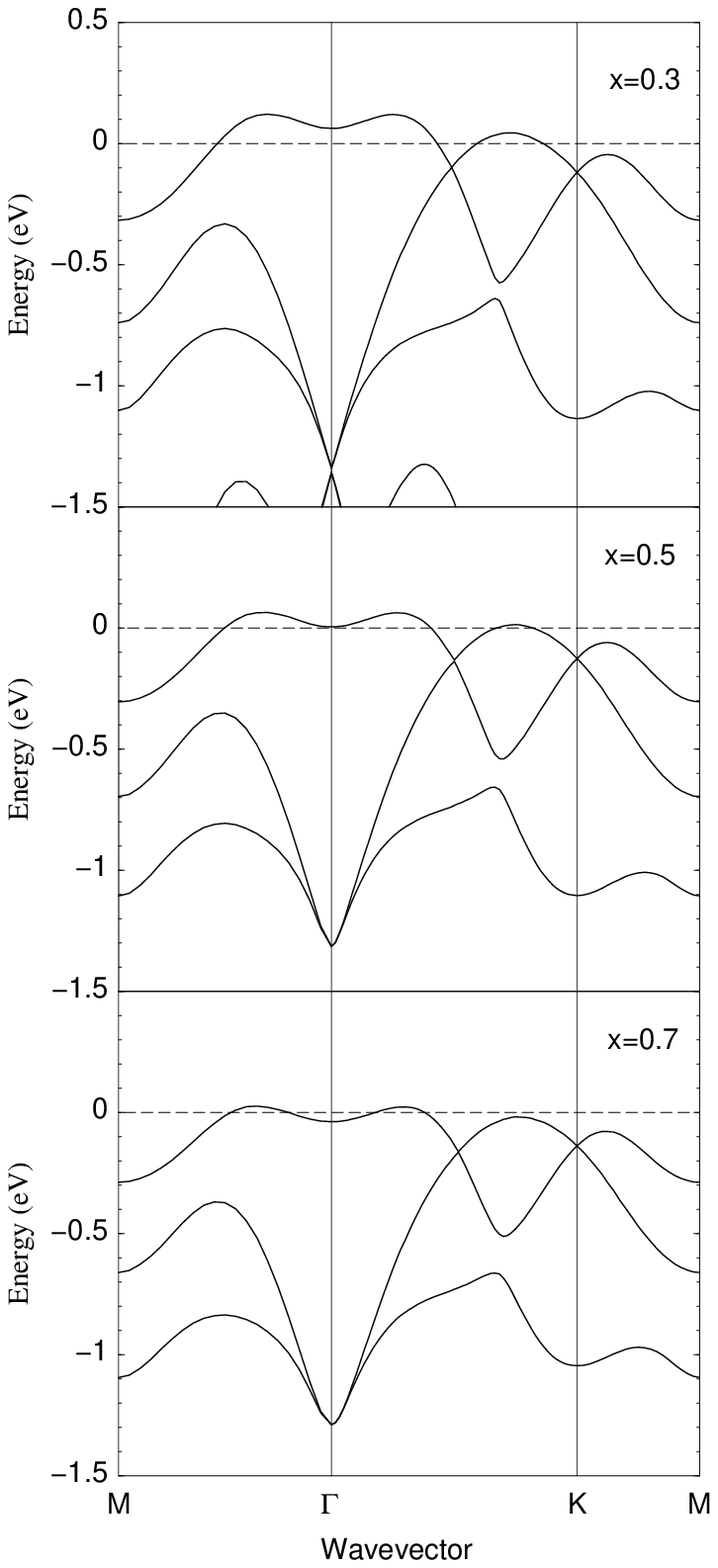}
\caption{LDA band structure of (CoO$_2$)$^{x-}$. Only the states derived from  $t_{2g}$ triplet
are shown. The horizontal dashed line indicates the Fermi level.}
\end{figure}

\newpage
\begin{figure}[h]
\includegraphics[width=8cm]{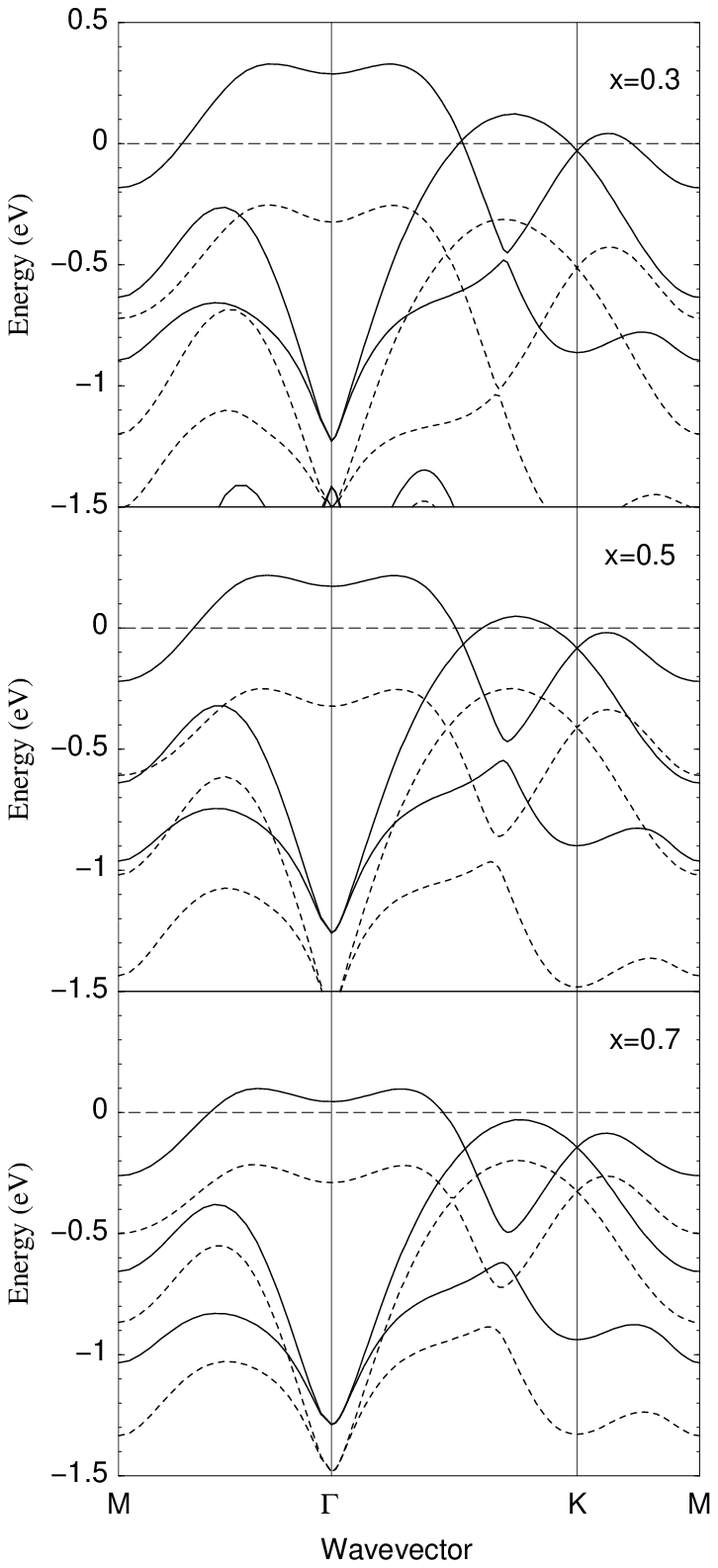}
\caption{LSDA band structure of (CoO$_2$)$^{x-}$ near the Fermi level.
Solid and dotted curves are minority and majority spin bands,
respectively. The horizontal dashed line indicates the Fermi level.}
\end{figure}
\newpage

\begin{figure}[h]
\includegraphics[width=8cm]{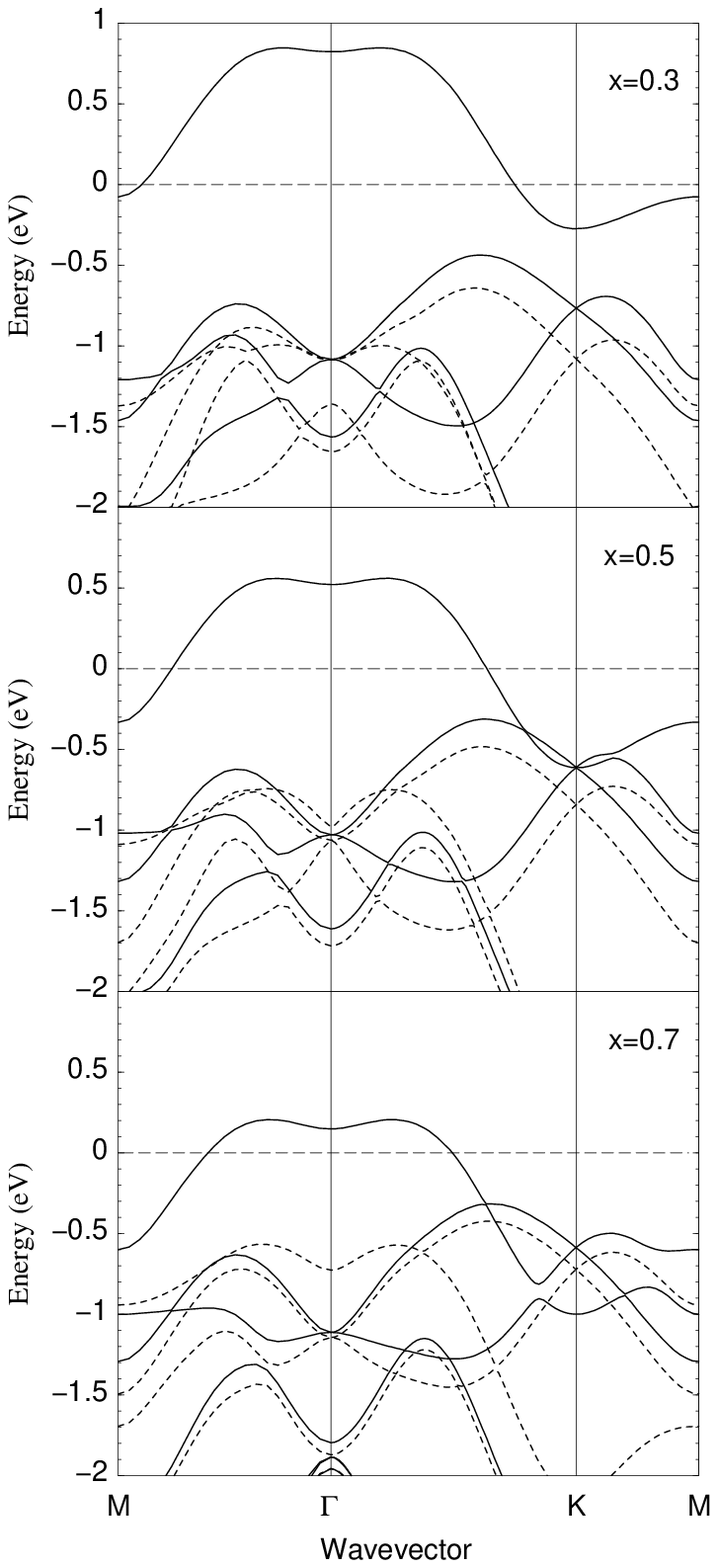}
\caption{LSDA+U band structure of (CoO$_2$)$^{x-}$ near the Fermi level.
Solid and dotted curves are minority  and majority spin bands,
respectively. The horizontal dashed line indicates the Fermi level.}
\end{figure}

\newpage
\begin{figure}[h]
\includegraphics[width=8cm]{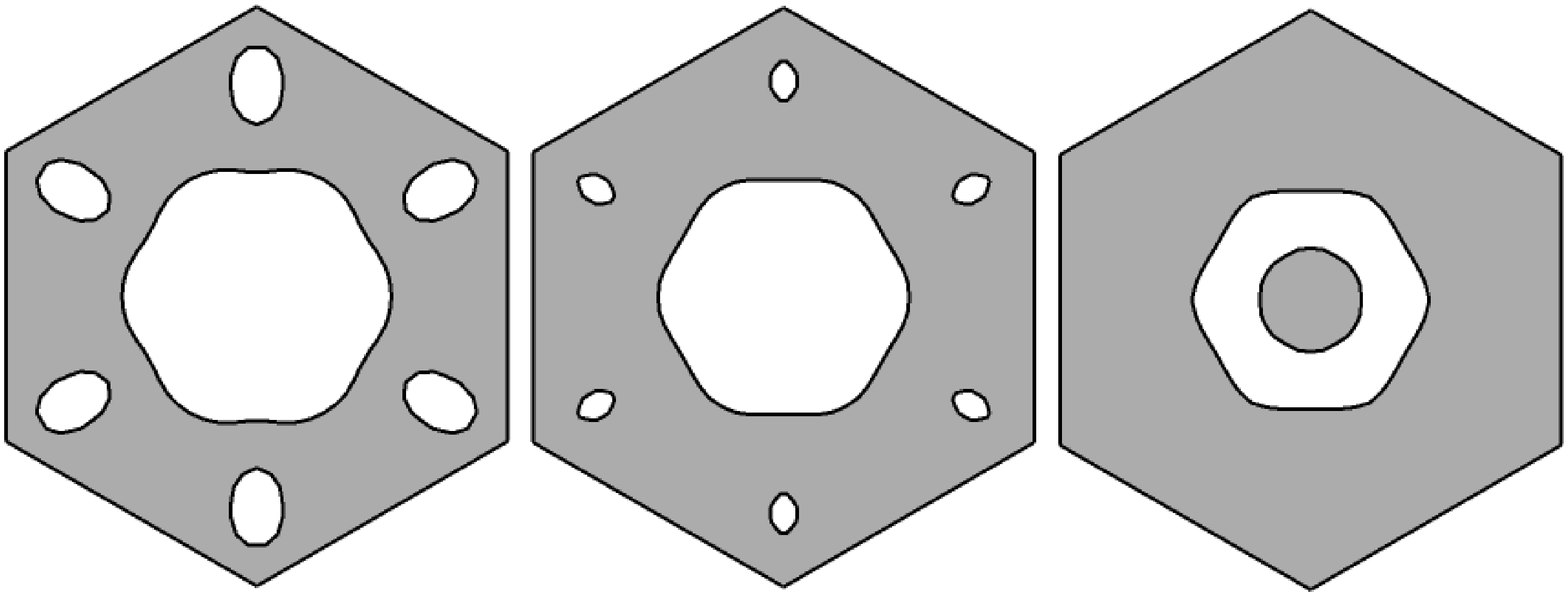}

\includegraphics[width=8cm]{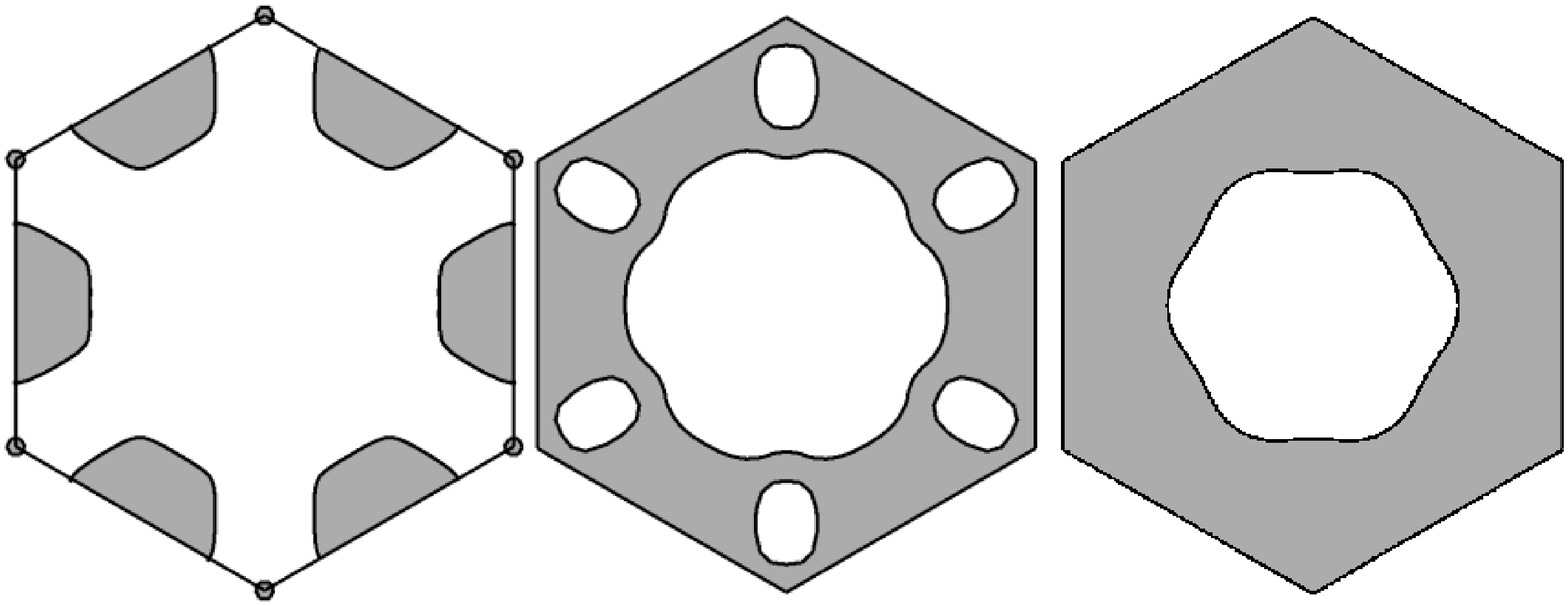}

\includegraphics[width=8cm]{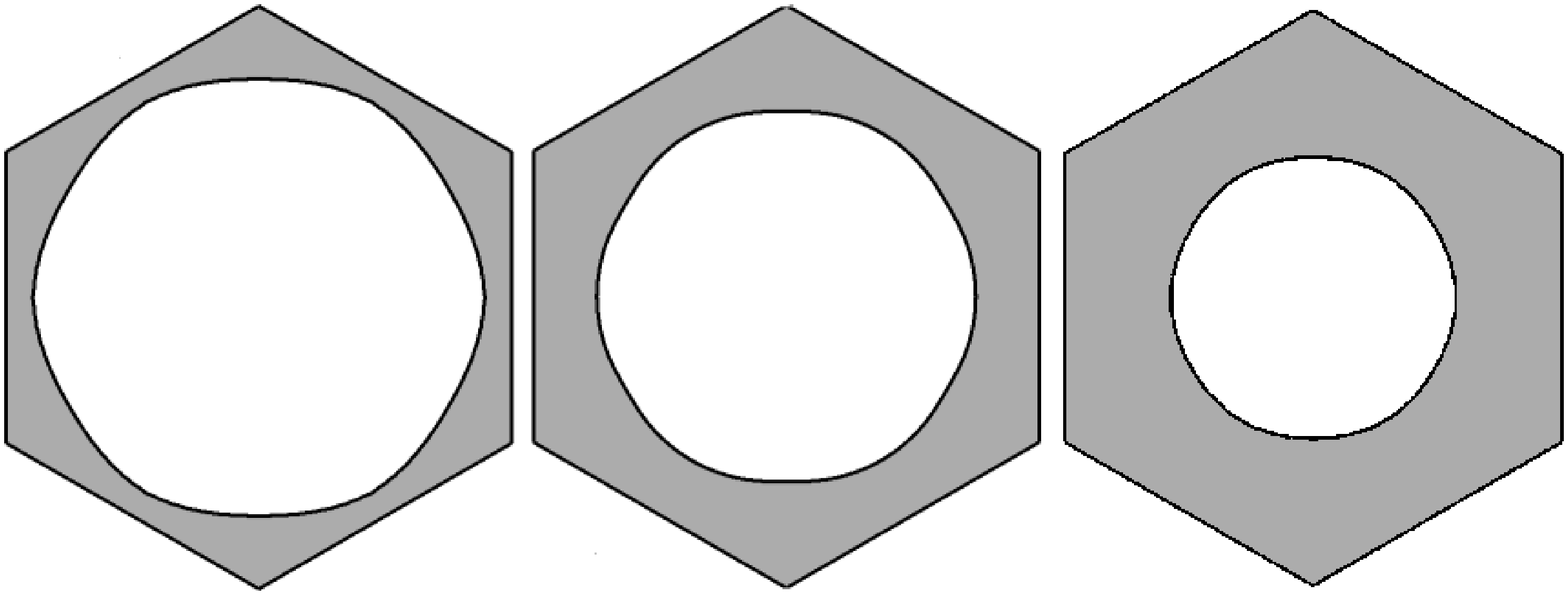}
\caption{Fermi surface topology of (CoO$_2$)$^{x-}$ calculated with
(a) LDA, (b) LSDA, and (c) LSDA+U for doping levels $x=$ 0.3, 0.5, and
0.7, from left to right, respectively. Shaded areas are electron states 
and empty areas are holes.}
\end{figure}

\begin{figure}[h]
\includegraphics[width=17cm]{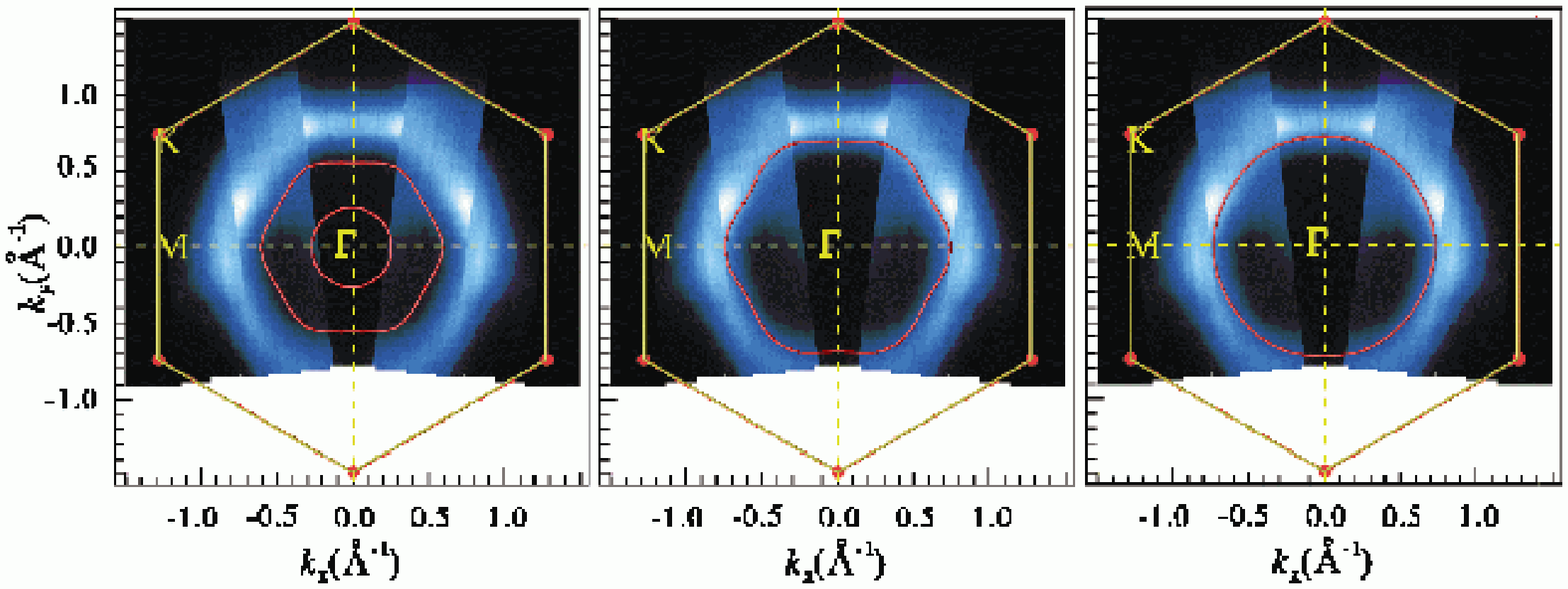}
\caption{Comparison between the measured and the calculated Fermi surface (red curve) of Na$_{0.7}$CoO$_2$
using LDA (left), LSDA (middle), and LSDA+U (right). The ARPES experimental result is taken
from the work of Hasan {\it et al..}\cite{Hasan04}}
\end{figure}


\begin{references}

\bibitem{supercond} K. Takada, H. Sakural, E. Takayama-Muromachi,
F. Izumi, R. A. Dilanian, and T. Sasaki, Nature {\bf 422} 53 (2003).



\bibitem{Lee04} K.-W. Lee, J. Kunes, and W. E. Pickett,
Phys. Rev. B {\bf 70}, 045104 (2004)
\bibitem{Zhang04} P. Zhang, W. Luo, V. H. Crespi, M. L. Cohen, and S. G. Louie,
Phys. Rev. B {\bf 70}, 085108 (2004).
\bibitem{Marianetti04} C. A. Marianetti, G. Kotliar, and G. Ceder,
Phys. Rev. Lett. {\bf 92}, 196405 (2004).
\bibitem{Johannes04_1}  M. D. Johannes and D. J. Singh,
Phys. Rev. B {\bf 70}, 014507 (2004).
\bibitem{Li04} Z. Li, J. Yang, J. Hou, and Q. Zhu, cond-mat/0403727.
\bibitem{Johannes04_3}  M. D. Johannes, D. A. Papaconstantopoulos, D. J. Singh, and M. J. Mehl,
cond-mat/0408698.
\bibitem{Wu04} W. B. Wu, D. J. Huang, J. Okamoto, A. Tanaka, H. J. Lin,
F. C. Chou, A. Fujimori, and C. T. Chen, cond-mat/0408467.
\bibitem{Zou04} L. J. Zou, J. L. Wang, and Z. Zeng, Phys. Rev. B {\bf 69}, 132505 (2004).
\bibitem{NaCo2O4} D. J. Singh, Phys. Rev. B {\bf 61}, 13397 (2000).

\bibitem{Kuroki04} K. Kuroki, Y. Tanaka, and R. Arita, Phys. Rev. Lett. {\bf 93}, 077001 (2004).
\bibitem{Johannes04_2} M. D. Johannes, I. I. Mazin, D. J. Singh, and D. A. Papaconstantopoulos,
Phys. Rev. Lett. {\bf 93}, 097005 (2004).

\bibitem{Hasan04} 
M. Z. Hasan, Y.-D. Chuang, D. Qian, Y. W. Li, Y. Kong, A. Kuprin, A. V. Fedorov, R. Kimmerling, 
E. Rotenberg, K. Rossnagel, Z. Hussain, H. Koh, N. S. Rogado, M. L. Foo, and R. J. Cava,
Phys. Rev. Lett. {\bf 92}, 246402 (2004).
\bibitem{Yang04} H.-B. Yang, S.-C. Wang, A. K. P. Sekharan, H. Matsui, S. Souma, 
T. Sato, T. Takahashi, T. Takeuchi, J. C. Campuzano, R. Jin, B. C. Sales, 
D. Mandrus, Z. Wang, and H. Ding, Phys. Rev. Lett. {\bf 92}, 246403 (2004).
\bibitem{LDA+U} V. I. Anisimov, J. Zaanen, and O. K. Andersen,
Phys. Rev. B {\bf 44}, 943 (1991).

\bibitem{magnetism1} T. Motohashi, R. Ueda, E. Naujalis, T. Tojo, I. Terasaki, T. Atake,
M. Karppinen, and H. Yamauchi, Phys. Rev. B {\bf 67}, 064406 (2003).
\bibitem{magnetism2} J. Sugiyama, H. Itahara, J. H. Brewer, E. J. Ansaldo,
T. Motohashi, M. Karppinen, and H. Yamauchi, Phys. Rev. B {\bf 67}, 214420 (2003).

\bibitem{Foo04} M. L. Foo, Y. Wang, S. Watauchi, H. W. Zandbergen, T. He,
R. J. Cava, and N. P. Ong, Phys. Rev. Lett. {\bf 92}, 248001 (2004).
\bibitem{Mukhamedshin04} I. R. Mukhamedshin, H. Alloul, G. Collin, and
N. Blanchard, cond-mat/0402074.
\bibitem{Huang04} Q. Huang, B. Khaykovich, F. C. Chou, J. H. Cho, J. W. Lynn, and Y. S. Lee,
cond-mat/0405677.
\bibitem{Wang04} N. L. Wang, D. Wu, G. Li, X. H. Chen, C. H. Wang, and X. G. Luo,
cond-mat/0405218.
\bibitem{Lee042} K.-W. Lee, J. Kunes, P. Novak, and W. E. Pickett,
cond-mat/0408411.

\end{references}
\end{document}